\documentclass[11pt]{article}
 \usepackage{epsf}
\newcommand{\eq }{\begin{equation}}
\newcommand{\en}{\end{equation}}
\newcommand{\eqa}{\begin{eqnarray}}
\newcommand{\ena}{\end{eqnarray}}

\newcommand{\ZZ}{\hbox{{\rm Z{\hbox to 3pt{\hss\rm Z}}}}}
\newcommand{\bra}{\langle}
\newcommand{\ket}{\rangle}

\newcommand{\oto}{\leftrightarrow}

\newcommand{\NP}[1]{Nucl.\ Phys.\ {\bf #1}}

\newcommand{\PR}[1]{Phys.\ Rev.\ {\bf #1}}

\newcommand{\MPL}[1]{Mod.\ Phys.\ Lett.\ {\bf #1}}

\begin{document}

\setcounter{footnote}{0}
\begin{titlepage}
\vskip0.5cm
\begin{flushright}
%{\tt draft}
DFTT 25/00\\
%{\tt draft}
\end{flushright}
\vskip0.5cm
\begin{center}
{\Large\bf Thermal operators and  cluster topology\\
in q-state Potts Model }
\end{center}
\vskip 1.3cm
\centerline{
M. Caselle${\small~^{a,b}}$, F. Gliozzi${\small~^{a,b}}$ and
S.Necco${\small~^c}$ }
 \vskip 1.0cm
 \centerline{\sl {\small$~^a$}   Dipartimento di Fisica
 Teorica dell'Universit\`a di Torino}
 \centerline{\sl {\small$~^b$}  Istituto Nazionale di Fisica Nucleare, Sezione di Torino}
 \centerline{\sl via P.Giuria 1, I-10125 Torino, Italy}
 \centerline{ \sl {\small$~^c$} DESY-IfH Zeuthen, Platanenallee 6,
  D-15738 Zeuthen, Germany}
 \vskip .4 cm

\begin{abstract}
We discuss a new class of identities  between correlation functions which arise
from a local $\ZZ_2$ invariance of the partition function of the $q-$state
Potts model on general graphs or lattices.  Their common feature is to relate the thermal
operators of the Potts model to some  topological properties of the Fortuin-Kasteleyn
clusters. In particular it turns out that any even correlation function can be
expressed in terms of observables which probe the linking properties of these clusters.
This generalises a class of analogous relations recently found in the Ising model.
\end{abstract}
\end{titlepage}
\section{Introduction}
One of the simplest and more studied models of statistical mechanics
is the $q-$state Potts model \cite{potts,wu}, which is the basic system
symmetric under the permutation group of $q$ elements. It can be
defined on any connected graph $G$ associating with each vertex
$i=1,2,\dots, v(G)$ the
spin variable $\sigma_i=1,2,\dots,q$ . Its  partition function at temperature
$T=1/\beta$ is taken to be
\eq
Z_G=\sum_{\{\sigma\}}e^{-\beta {\cal H}}
\label{zpotts}
\en
where
${\cal H} = -\sum_{\bra ij \ket} \delta_{\sigma_i \sigma_j}$
 with $\bra ij \ket$ ranging over pairs of adjacent
vertices in $G$.

A long-standing problem is how to characterise geometrically in 
this kind of models the fluctuations near a critical point. Contrary
to the naive expectation, the clusters made of adjacent sites with aligned
spins do not play an important role in this respect. A different 
definition of cluster was proposed for the Ising model \cite{ck} and
generalised to the $q-$state Potts model \cite{cp}. These clusters are
defined  as adjacent sites with the same spin connected by bonds  with
probability $p=1-e^{-\beta}$. Within such a definition, these clusters
behave correctly at the critical point, in the sense that their radius
and the density of the percolating cluster scale with the correct
critical exponents. 
  
The partition function (\ref{zpotts}) can be rewritten 
in terms of these clusters using the Fortuin Kasteleyn representation: 
\cite{whit,kf}
\eq
Z_G(q,v) = \sum_{G^\prime \subseteq G}v^{e(G^\prime)} q^{k(G^\prime)},
\label{zqv}
\en
where  $ v=\frac p{1-p}=e^\beta-1$; the summation is over all 
spanning  subgraphs of $G$, namely the
subgraphs made with all the vertices of $G$ {\sl i.e.} with 
$v(G^\prime)=v(G)$;  $e(G^\prime)$ is the
number of edges of $G'$, called {\sl active} bonds, and $k(G')$ the 
number of connected components or Fortuin-Kasteleyn (FK) clusters.
This formulation of the partition function, sometimes called a
di-chromatic polynomial, enables one to generalise $q$ from
positive integers to real and complex values. In particular $q=0$
corresponds to the tree percolation problem and $q=1$ is the random 
percolation problem.
\vskip .2 cm

In the present work we establish some new  exact topological properties of
these clusters. They arise from a class of identities  between
correlation functions which are valid for any graph  $G$ and come
from the invariance of the partition function (\ref{zpotts}) with respect to a
local $\ZZ_2$ transformation. Their origin suggests to call them
Ward identities (see Sect. 2). Their common feature is to relate the thermal
operators of the Potts model to some  linking properties of the FK
 clusters and generalise a class of analogous relations 
recently found in the Ising model \cite{cg}.

 It turns out that any even correlation function can be
 expressed in terms of a new kind of observables, which 
depend only on the closed circuits of these clusters (see Sect. 3).
This suggests new, powerful methods to evaluate even correlators 
which can be easily implemented in numerical experiments: let
$G'_1,G'_2,\dots,G'_i,\dots$ be a sequence of spanning subgraphs
in statistical equilibrium with the partition function (\ref{zqv}),
generated by any updating algorithm. Deleting all the edges which do
 not belong to any circuit of $G'_i$ produces a dramatic
 simplification of the configuration with no information loss on 
thermal properties of the system. In the simplified configurations 
the evaluation of the new kind of observables is a much easier task.
This leads to improved estimators of the thermal operators of this model.  
   
\vskip .2 cm
\section{Ward Identities}
An elementary derivation of the simplest of these identities is the
following.  Let  $\ell$ be any edge of $G$. Denote by $G_\ell^+$ the spanning
subgraphs containing $\ell$  and by $G_\ell^-$ the
others. Clearly the set $\{G'\}$ of all the spanning subgraphs may  be written
as the sum of two disjoint subsets $ \{G'\}=\{G_\ell^+\}\cup\{G_\ell^-\}$.
Denote by $\delta_\ell$ the   $\ZZ_2$
transformation on $Z_G(q,v)$ which
deletes the edge $\ell$
from any subgraph of type $G_\ell^+$ and adds it to any subgraph
of type $G_\ell^-$. This is a symmetry of the partition function, namely,
\eq
\delta_\ell\left[Z_G(q,v)\right]=Z_G(q,v)\;\;,
\label{della}
\en
since $\sum_{G^\prime \subseteq G}=\sum_{G_\ell^+ \subseteq G}+\sum_{G_\ell^- \subseteq G}$
and $\delta_\ell$ exchanges the two subsets $\{G_\ell^+\}\oto\{G_\ell^-\}$.
On the contrary the two
partial sums are not invariant, but their way of
transforming under $\delta_\ell$ can be explicitly evaluated. We have, of course,
\eq
\delta_\ell\left[ e(G_\ell^\pm)\right]=e(G_\ell^\pm)\mp 1\;\;.
\label{delle}
\en
The action of $\delta_\ell$ on the number of clusters $k(G^+_\ell)$
depends on a  topological property of $\ell$: if it belongs to
a circuit of $G_\ell^+$,  $k$ is kept invariant
\eq
\delta_\ell\left[ k(G_\ell^+)\right]=k(G_\ell^+)\;\;;
\label{black}
\en
we call black bonds the edges with this property \cite{cg}. If $\ell$
do not belongs to any closed circuit of $G_\ell^+$  we have
\eq
\delta_\ell \left[k(G_\ell^+)\right]=k(G_\ell^+)+1\;\;,
\label{gray}
\en
and  call it  grey bond or  bridge.
Using (\ref{delle}), (\ref{black}) and (\ref{gray}) we obtain
 \footnote{This way of reasoning is very similar to the
one used to describe the basic properties of the Tutte polynomial of a
graph \cite{tutte}, which is a simple variant of the partition
function (\ref{zqv}); see also \cite{bollo}.}
\eq
\delta_\ell\left[\sum_{G_\ell^+ \subseteq G}v^{e(G_\ell^+)}q^{k(G_\ell^+)}\right]=
\sum_{G_\ell^+ \subseteq G}[\frac 1v \pi_{G_\ell^+}(\ell)+
\frac qv(1-\pi_{G_\ell^+}(\ell))]
v^{e(G_\ell^+)}q^{k(G_\ell^+)},
\label{plus}
\en
where $\pi_{G'}(\ell)$ denotes a projector on the black bonds defined
as follows
\eq
\pi_{G'}(\ell)=\cases{1& if $\ell$ is a black bond\cr
0& otherwise}
\en
For the other partial sum we can write, more simply,
\eq
\delta_\ell\left[\sum_{G_\ell^- \subseteq G}v^{e(G_\ell^-)}q^{k(G_\ell^-)}\right]=
\sum_{G_\ell^+ \subseteq G}v^{e(G_\ell^+)}q^{k(G_\ell^+)}\;\;.
\label{minus}
\en
From (\ref{della}), (\ref{plus}) and (\ref{minus}) one obtains
immediately
\eq
1=\frac{v+q}v\bra \rho(\ell)\ket-\frac{q-1}v\bra \pi(\ell)\ket~,
\en
where $\bra\rho(\ell)\ket=\sum_{G_\ell^+}P(G_\ell^+)$, with $P(G')=v^{e(G')}q^{k(G')}/Z_G(q,v)$.

Summing over all the edges $\ell=1,2\dots n\equiv e(G)$ yields
\eq
 n=\frac{v+q}v\bra e\ket-\frac{q-1}v\bra b\ket~~,
\label{ward1}
\en
where
$\bra e\ket=\sum_{G'}e(G')P(G')$ is the mean number of
 edges of $G'$, called active bonds and similarly
$\bra b\ket= \sum_{G' }b(G')P(G')$,
where $b(G')$ is the  number of black bonds 
in the subgraph $G'$. This is the first
Ward identity. Since $\bra e\ket$ is related to the internal
energy $E$ according to
\eq
E\equiv-\frac{\partial\log Z_G}{\partial\beta}=
-\frac{v+1}v\bra e\ket~~,
\label{energy}
\en
we can rewrite the above identity in the form
\eq
-E=\frac{v+1}{v+q}e(G)+\frac{(q-1)(v+1)}{v(v+q)}\bra b\ket~~,
\label{enerb}
\en
which gives the internal energy $E$ in terms of black bonds. This
illustrates a common feature of this kind of Ward identities: all the
thermal properties  of the model are encoded in the black bond configurations,
as we shall see shortly. The fact that the energy counts this special
kind of bonds has been already observed some time ago \cite{cardy}.

As a little diversion, let us remark that if $G$ were a tree $T$,
then $b(T)\equiv b(G')\equiv0$. This leads to $E= -\frac{v+1}{v+q}e(T)$,
which is a well known result of the $q-$state Potts model on any tree with
sub-exponential growth \cite{bc}; it also holds in the symmetric phase of any  Bethe
lattice .

It is also interesting to study Eq. (\ref{ward1}) in the pure percolation
limit $q\to 1$. Taking this limit together with the Taylor expansion
of $\bra e\ket$
\eq
\bra e\ket=\frac {v\, e(G)}{v+1} +
(q-1)\left[\bra e\, k\ket -
\bra e\ket\bra k\ket\right]_{q=1} +\dots,
\en
yields the curious identity
\eq
\frac{\bra e-b\ket_{q=1}}{v+1}=
\left[ \bra e\ket\,\bra k\ket-\bra
e\,k\ket\right]_{q=1}~~,
\en
relating the mean number of bridges to the connected correlator
between edges and clusters.
\vskip .2cm
Other Ward identities can be simply obtained by applying the above
$\ZZ_2$ transformation to more than one edge. In particular, the
two-edge transformation leads to  the following expression of the
specific heat $C=\frac{\partial E}{\partial T}$ in terms of black
bonds
\begin{eqnarray}
\frac C{ \beta^2}+E+\frac{E^2}{e(G)}&=&
\frac{(q-1)(v+1)^2}{v(v+q)}{\Big\{}
2\bra h\ket+
\nonumber
\\
&+&(q-1)\Big[\bra b^2\ket-\bra b\ket^2-
\frac{\bra b\ket}n\left(n-\bra
b\ket\right)\Big]{\Big\}}
\label{heat}
\end{eqnarray}
with $\bra h\ket=\sum_{G'}h(G')P(G')$, where $h(G')$ denotes the number of the
cutting pairs of the spanning subgraph $G'$, {\sl i.e.} the pairs of black
bonds such that their deletion increases the number of clusters.

\section{Cutting the bridges}
We come now to the other purpose of this paper, namely
to show that not only the internal energy or the specific heat, but
also any correlation function involving an even number of sites my be evaluated
in terms of black bond configurations $B(G')$, {\sl i.e.} the graphs
generated by  deletion of all the bridges in the spanning subgraphs.
To this end it is convenient to slightly generalise the model in the following
way: we assign to any oriented edge $\bra ij\ket$ of $G$ a twist, namely
an integer $\tau_{ij}=0,\pm1,\dots,\pm(q-1)$ and define a 
modified Hamiltonian
\eq
{\cal H}_\tau = -\sum_{\bra ij \ket}
\delta_{\sigma_i, \sigma_j+\tau_{ij}}~~,
\en
where the sum $\sigma_i+\tau_{ij}$ is taken modulo $q$. 
An edge $\bra ij\ket$ with
$\tau_{ij}\not=0$ is said to be twisted. 
In order to get the Fortuin Kasteleyn representation of the twisted
partition function $Z_{G,\tau}=\sum_{\{\sigma\}}{\rm e}^{-\beta{\cal
    H}_\tau}$,
we can rewrite it as
\eq
Z_{G,\tau}=\sum_{\{\sigma\}}\prod_{\bra ij\ket}(1+v
\delta_{\sigma_i,\sigma_j+\tau_{ij}})~,~~(v={\rm e }^\beta-1)~,
\en
and expand the product as a sum of terms corresponding to the possible
choices of summand from each factor. Each edge of the graph $G$ can
appear with a factor 1 or $ v\delta_{\sigma_i,\sigma_j+\tau_{ij}}$; in
the latter case the edge is said to be an active bond. In a term with 
$e(G')$ active bonds we have a factor $v^{e(G')}$ together with a
product of delta functions which forces  to zero every configuration 
$G'$ in which all the spins connected by the active bonds do not satisfy 
the constraint $\delta=1$. If there are no twists in $G'$ this
constraint implies that all the spins of each cluster are aligned and
the total number of spin configurations in a situation in which $G'$
is formed by $k(G')$ clusters is $q^{k(G')}$. If, on the contrary,
there are twisted edges on $G'$, we get a non-vanishing contribution 
only if there are no frustrations, namely if and only if the algebraic
sum of the twists along {\sl any} circuit of $G'$ is zero modulo $q$.
In order to select these subgraphs let us  introduce a projector 
$\varpi_\tau(G')$ defined as 
\eq
\varpi_\tau(G')=\cases{1&if  $G'$ has no frustrations\cr
0& otherwise.}
\en
Note that this projector picks out a subgraph $G'$ regardless the
location of its bridges, thus it can only be a function of the black
bond configurations $B(G')$:
\eq
\varpi_{\tau}(G')= \varpi_{\tau}(B(G'))~.
\label{taub}
\en
In terms of this projector we have
\eq
Z_{G,\tau}\equiv\sum_{\{\sigma\}}e^{-\beta {\cal H}_\tau}=
\sum_{G'\subseteq G}\varpi_{\tau}v^{e(G')}
q^{k(G')}~;
\label{ztqv}
\en
this leads to the fundamental identity
\eq
\frac{Z_{G,\tau}}{Z_G}=\bra\varpi_\tau\ket~~,
\label{main}
\en
where the expectation value is taken with respect to the standard,
untwisted Hamiltonian. This defines a (complete) set of physical observables,
one for each choice of $\tau$, which depend only on the black bond
configurations and probe the linking properties of the FK clusters.

Thus the problem reduces to show that any even correlation function
can be expressed in terms of  suitable combinations of
$\varpi_\tau$'s. This can be done, at least in principle, by
writing the expectation value  (\ref{main}) in terms of spin variables
\eq
\bra\varpi_\tau\ket=
\frac{\sum_{\{\sigma\}}{\rm e}^{-\beta{\cal H}_\tau}}
{\sum_{\{\sigma\}}{\rm e}^{-\beta{\cal H}}}
=\bra {\rm e}^{\beta\sum_{\bra ij\ket}
( \delta_{\sigma_i,
\sigma_j+\tau_{ij}}-\delta_{\sigma_i\sigma_j})}\ket
\label{disorder}
\en
and then expanding the exponential. As examples, it is easily
verified that a single twisted edge generates the first Ward identity
(\ref{ward1}) and the two-edge twist yields eq.(\ref{heat}).

Note that eq. (\ref{disorder}) has the typical form of the expectation
value of correlators of disorder observables \cite{kc}. 
Measuring disorder observables is not an easy task: one needs
expectation values of operators which are exponentially suppressed 
on dominant configurations of the statistical ensemble. In particular,
the numerical evaluation of operators involving a large number of
twisted links represents a very hard sampling problem.
Actually Eq. (\ref{taub}) yields a simple, effective way out. Indeed
$\varpi_\tau(B(G'))$ is an improved estimator of the same observable.
 The evaluation
of $\varpi_\tau$ on the black bond configurations is much less noisy
 than in the spin
configurations, because in the former case $\varpi_\tau$ can only take
 the  values 0 and 1,
while in the latter case the exponential of spin variables has a much
 larger variance,
especially for large distance correlators. As a matter of fact,
 powerful algorithms to evaluate these new estimators have been
 constructed for
 the surface tension in 3D Ising model \cite{h93} and for the  Wilson
 loops  Polyakov correlators
 in 3D $\ZZ_2$ gauge model \cite{cfghp}-\cite{gp}. Their extension to
 a general $q-$ state Potts model is now straightforward.

\vskip .2 cm
This work has been sup\-ported in part by the Minis\-te\-ro ita\-lia\-no
dell'Univer\-sit\`a e del\-la Ricerca Scientifica e Tecnologica.

\end{document}